\documentclass[aps,twocolumn,amsmath,amssymb, superscriptaddress]{revtex4}
\usepackage[pdftex]{graphicx}
 \usepackage{amsmath}
\usepackage[T1]{fontenc}
\usepackage{amssymb}
\usepackage{amsfonts}
\usepackage{bm}
 \usepackage{amsmath} 
\usepackage{lipsum}
\usepackage{amsfonts} 
\usepackage{amssymb, mathrsfs}
\usepackage{braket}
\usepackage{graphicx} 
\usepackage{subfigure}
\usepackage{bbm}
\def\beq{\begin{equation}}
\def\eeq{\end{equation}}
\def\bsp{\begin{split}}
\def\esp{\end{split}}
\def\bea{\begin{eqnarray}}
\def\eea{\end{eqnarray}}
\def\ba{\begin{array}}
\def\ea{\end{array}}

\def\dg{\dagger}

\def\lb{\left(}
\def\rb{\right)}

\def\l.{\left.}
\def\r.{\right.}

\def\ra{\rangle}
\def\la{\langle}

\def\bo{{\vec k}}

\begin{document}

\date{\today}
\title{Topological  thermal Hall effect due to Weyl magnons}
\author{S. A. Owerre}
\affiliation{Perimeter Institute for Theoretical Physics, 31 Caroline St. N., Waterloo, Ontario N2L 2Y5, Canada.}

\begin{abstract}
 We present the first theoretical evidence of  zero magnetic field topological (anomalous) thermal Hall effect  due to Weyl magnons. Here, we consider Weyl magnons in   stacked noncoplanar frustrated kagom\'e antiferromagnets recently proposed by  Owerre,  [arXiv:1708.04240]. The Weyl magnons in this system result from  macroscopically broken  time-reversal symmetry by the scalar spin chirality of noncoplanar chiral spin textures. Most importantly, they come from the lowest excitation, therefore they can be easily observed experimentally at low temperatures due to the population effect.  Similar to electronic Weyl nodes close to the Fermi energy, Weyl magnon nodes in the lowest excitation are the most important. Indeed,  we show that the topological (anomalous) thermal Hall effect in this system arises from nonvanishing Berry curvature due to Weyl magnon nodes in the lowest excitation, and it depends on their distribution (distance) in momentum space.   The present result paves the way to directly probe low excitation Weyl magnons and macroscopically broken time-reversal symmetry  in three-dimensional frustrated magnets with the anomalous thermal Hall effect. 
 \end{abstract}
\maketitle
\section{Introduction}
 In the classical Hall effect \cite{Hall}, a magnetic field is applied perpendicular to the direction of an electric current  in  metals  and causes charge carriers to  experience  a Lorentz force. The propagation of electric current is deflected in circular orbits by the Lorentz force and charges accumulate on the edge of the material which causes a voltage difference called the Hall voltage.  Quantum mechanically,  the circular orbits can be quantized as Landau levels and give rise to a quantized Hall conductivity termed the integer quantum Hall effect \cite{kli,thou,lau1,nov, tsu,thou1}. The  anomalous Hall effect arises from the quantum Berry curvature due to nontrivial electronic band topology \cite{nag}. It can manifest spontaneously and scales with the magnetization in ferromagnets. Recently, the anomalous Hall effect has been reported in zero magnetic-field antiferromagnets with  strong  spin-orbit coupling (SOC)  and vanishingly small magnetization \cite{chen,nay,nak,kub, ito, kiy, ito1,suz}. It can also manifest  as a topological Hall effect  resulting from nontrivial noncoplanar chiral spin textures \cite{ele,ele1,ele0, ohg, sur} even in the absence of SOC.  Its quantized form is termed the quantum anomalous Hall effect \cite{fdm,ele2,cha}.

The anomalous thermal Hall effect also requires the presence of a Berry curvature similar to the anomalous Hall effect, but in this case a transverse heat current flows under the influence of a longitudinal thermal gradient.   In contrast to electronic charged particles,   the carriers of the anomalous thermal Hall effect  in magnetic systems are charge-neutral bosonic quasiparticles such as magnons, phonons, triplons, and spinons, and they do not experience a Lorentz force, but the Berry curvature can be regarded as an effective magnetic field in momentum space.   In insulating quantum ferromagnets the presence of spontaneous magnetization together with an out-of-plane Dzyaloshinskii-Moriya  (DM) interaction \cite{dm,dm2}  (or SOC)   in the direction of the magnetization breaks time-reversal ($\mathcal T$) symmetry  and generates a nonzero Berry curvature \cite{shin1,alex0,alex2,alex4,alex44, alex2a}, which induces  an anomalous thermal Hall effect \cite{shin1,alex0,alex2,alex4,alex44,alex1, alex1a, alex6, alex2a}.   This effect  has also been observed in frustrated magnets with spin liquid states  \cite{wat, hir}, and recently in multiferroics \cite{ide}.   To date, however,  the anomalous thermal Hall effect has not been observed experimentally in magnetically ordered insulating antiferromagnets with vanishingly small magnetization. To our knowledge, there is no theoretical or experimental study of the anomalous thermal Hall effect induced by Weyl magnons (WMs).

The insulating quantum antiferromagnets are of great interest and they behave differently from insulating quantum ferromagnets. For example, in   frustrated kagom\'e antiferromagnets  with only an out-of-plane DM interaction, a  conventional non-collinear $120^\circ$  magnetic structure with zero scalar spin chirality can be induced.  It possesses an effective $\mathcal T$ symmetry and leads to vanishing Berry curvature, thus forbids an anomalous thermal Hall effect. This scenario is strikingly different from ferromagnets  where the out-of-plane DM interaction inevitably leads to an anomalous thermal Hall effect \cite{shin1,alex0,alex2,alex4,alex44,alex1, alex1a, alex6}.  Interestingly, most frustrated  kagom\'e antiferromagnets intrinsically possess  both  in-plane and out-of-plane DM interactions. While the out-of-plane DM component  stabilizes  a non-collinear $120^\circ$  spin structure, the in-plane DM component  induces canting out-of-plane and leads to  a noncoplanar chiral spin structure  with nonzero scalar spin chirality, which breaks $\mathcal T$ symmetry macroscopically \cite{men1, sup1a,zhe,zhe1,han,han1, sch}. A non-negligible interlayer coupling establishes a three-dimensional (3D) spin structure. Therefore, we expect that WMs and the associated anomalous thermal Hall effect  should exist in stacked  frustrated kagom\'e antiferromagnets even with vanishingly small magnetization. 

 In this paper, we show that the topological (anomalous) thermal Hall effect is indeed present  in stacked frustrated  kagom\'e antiferromagnets  with WM nodes.   The present author  had proposed WMs in this system, albeit  at zero in-plane DM interaction and nonzero magnetic field \cite{sol}. In the current study, we consider the effects of both the in-plane and out-of-plane DM interactions at zero magnetic field. The presence of an in-plane DM interaction  at zero magnetic field  provides an  intrinsic property of the WMs in this system, which leads to intrinsic topological (anomalous) thermal Hall effect.   Moreover, the WMs in this system  are different from those of pyrochlore antiferromagnets \cite{fei,new} by the presence of the scalar spin chirality of noncoplanar chiral spin texture. Therefore, they are provided by explicit macroscopically broken $\mathcal T$ symmetry as opposed to  implicit broken $\mathcal T$ symmetry by the magnetic order in pyrochlore antiferromagnets.  
 
   Another issue with WMs in pyrochlore antiferromagnets \cite{fei,new} is that they appear in the absence of the DM interaction, however gapped topological magnon bands were recently found in pyrochlore antiferromagnets with DM interaction \cite{lau}, which suggests that the WMs in pyrochlore antiferromagnets could be gapped out by the DM interaction.  In addition,  the WMs in pyrochlore (anti)ferromagnets \cite{mok,su,fei} occur above the lowest excitation at high energy. However, in any bosonic system the lowest excitation is thermally populated at low temperatures due to the Bose function, and makes dominant contributions to the  thermal Hall conductivity \cite{alex0,alex2, alex2a}.   Hence, due to the population effect the WMs in pyrochlore (anti)ferromagnets \cite{mok,su,fei} will not contribute to the anomalous thermal Hall effect at low temperatures. Therefore, previous  experimentally reported thermal Hall conductivity in pyrochlore ferromagnets  \cite{alex1,alex1a} and a subsequent theoretical calculation \cite{alex2a} are definitely not a consequence of  recently proposed WMs in this system \cite{mok,su}.
 
The most important property of WMs in the current model is that they come from the lowest magnon excitation, hence they contribute significantly to the thermal Hall conductivity at low temperatures.  In this regard,  it is valid to say that the most important WM nodes with potential applications are definitely those in the lowest excitation \cite{foot}.  We show that  the topological (anomalous) thermal Hall effect of WMs in this system depends on the  distribution  and distance between the WM nodes in momentum space. In general, the thermal Hall conductivity is a 3-pseudo-vector $\lb \kappa_{yz}^x, \kappa_{zx}^y, \kappa_{xy}^z\rb$, however the first two components vanish because  the distribution of the WM nodes in momentum space comes in pairs of opposite chiralities and leads to zero net Berry curvature in the plane perpendicular to the $k_x$ or $k_y$ momentum direction. But the plane perpendicular to the out-of-plane $k_z$ momentum direction contains WMs with nonzero net Berry curvature across the planes, which yields a finite $\kappa_{xy}^z$.  The 3D thermal Hall conductivity  can be separated as
\begin{align}
\kappa_{xy}^z= \int_{-\pi}^{\pi} \frac{d k_z}{2\pi}\kappa_{xy}^{\text{2D}}(k_z),
\label{ATHC}
\end{align}
  where $\kappa_{xy}^{\text{2D}}(k_z)$ is a set of 2D thermal Hall conductivity in the $k_x$-$k_y$ plane \cite{shin1} parameterized by $k_z$. It is given by

\begin{align}
\kappa_{xy}^{\text{2D}}(k_z)=- T\int \frac{\text{d}\bo_\parallel}{(2\pi)^2}\sum_{n=1}^N c_2\big( f^B[E_n(\bo_\parallel,k_z)]\big )\Omega_{n\bo_\parallel}^z(k_z),
\label{ATHC1}
\end{align}
 where $\bo_\parallel=(k_x,k_y)$, $T$ is the temperature, $c_2$ is a function of the Bose function and $\Omega_{n\bo_\parallel}^z(k_z)$ is the Berry curvature. 
 
 Since the Berry curvature is dominant  near the WM nodes, the major contribution  to the thermal Hall conductivity comes from these nodes at the lowest magnon band due to the Bose function. Hence, at  $T\neq 0$ the  empirical expression for the topological (anomalous) thermal Hall conductivity can be written as
 \bea 
 \kappa_{xy}^z\propto \sum \Delta k_0^i,
 \label{sep}
 \eea  
 where $\Delta k_0^i$ is the separation of the WM nodes along the $k_z$ momentum direction, which depends on the scalar spin chirality of noncoplanar chiral spin texture induced by the in-plane DM interaction. This  relation is akin to the anomalous Hall conductivity in electronic Weyl semimetal \cite{ya,bur}.  Indeed, when the WM nodes annihilate at the Brillouin zone (BZ) boundary, the system becomes a  fully gapped 3D topological magnon insulator with similar features to 2D counterparts \cite{zhh,mhe,alex6a,sol1}. 

\section{Zero magnetic field spin model}
\begin{figure}
\includegraphics[width=1\linewidth]{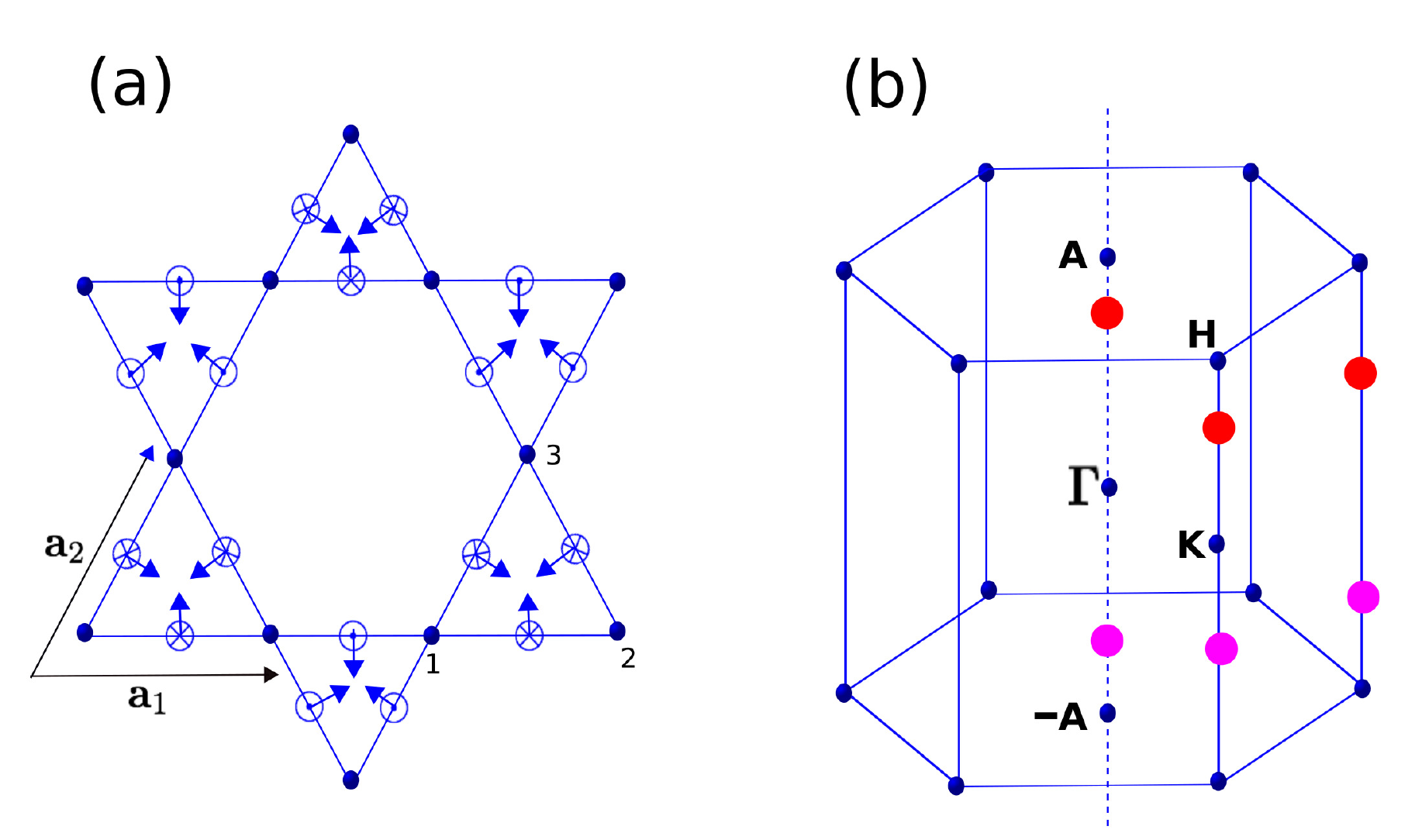}
\caption{Color online. (a). Top view of kagom\'e layers stacked perfectly along the (001) direction. It contains  in-plane (arrows) and out-of-plane (dotted and crossed circles) DM interactions.  The numbers ($1,2,3$) label the sublattices.  The in-plane unit vectors are ${\bf a}_1=(1,0,0)$ and ${\bf a}_2=(1/2,\sqrt{3}/2,0)$. The unit vector along the stacking direction ${\bf a}_3=(0,0,1)$ is not depicted. (b). The 3D Brillouin zone. Red and pink dots denote the locations of the lowest excitation WM nodes with opposite chiralities along the $k_z$ momentum direction. }
\label{KL}
\end{figure}

In Ref. \cite{sol}, we studied Weyl magnons in the stacked kagom\'e antiferromagnets induced by an external magnetic field. In this paper, we consider the Weyl magnons and the associated topological thermal Hall effect at zero magnetic field, but with the inclusion of the in-plane DM interaction. In this case the Hamiltonian for stacked  kagom\'e antiferromagnets   is governed by 
\begin{align}
\mathcal H&=J\sum_{\la ij\ra,\ell} {\vec S}_{i,\ell}\cdot{\vec S}_{j,\ell}+\sum_{\la ij\ra,\ell}{\vec D}_{ij}\cdot {\vec S}_{i,\ell}\times{\vec S}_{j,\ell} \nonumber\\&+ J_c\sum_{i,\la \ell \ell^\prime\ra}{\vec S}_{i,\ell}\cdot{\vec S}_{i,\ell^\prime},
\label{mod}
\end{align}
where $i$ and $j$ denote nearest neighbour sites on  the kagom\'e  layers, $\ell$ and $\ell^\prime$ label the layers. The first term is the intralayer antiferromagnetic Heisenberg exchange interaction.  The second term is the DM interaction \cite{dm,dm2}   due to lack of inversion symmetry between two sites on each layer.  Using the convention in ref.~\onlinecite{men1} we take  ${\vec D}_{ij}=(D_p\cos\varphi,D_p\sin\varphi,D_z)$, where $\varphi$ is the angle between the projection of the spin and the kagom\'e plane.  The orientations of the in-plane and out-of-plane DM vectors $D_p$ and $D_z$ respectively are depicted in Fig.~\eqref{KL}. The out-of-plane DM component $D_z>0$ stabilizes  a $120^\circ$ non-collinear spin configuration with positive vector chirality and retain U(1) rotational invariance about the out-of-plane $z$ direction, whereas the in-plane DM component  $D_p$ induces canting out-of-plane and leads to a noncoplanar chiral spin structure with broken time-reversal and rotational symmetries \cite{men1, sup1a,zhe,zhe1,han,han1, sch}. The last term is an unshifted interlayer antiferromagnetic interaction that establishes a 3D spin structure.  The most important physical consequence of this model is that all the interactions are allowed in real kagom\'e antiferromagnetic materials. 
For $\varphi=0$ the classical ground state energy  is given by

 \begin{figure*}
\includegraphics[width=1\linewidth]{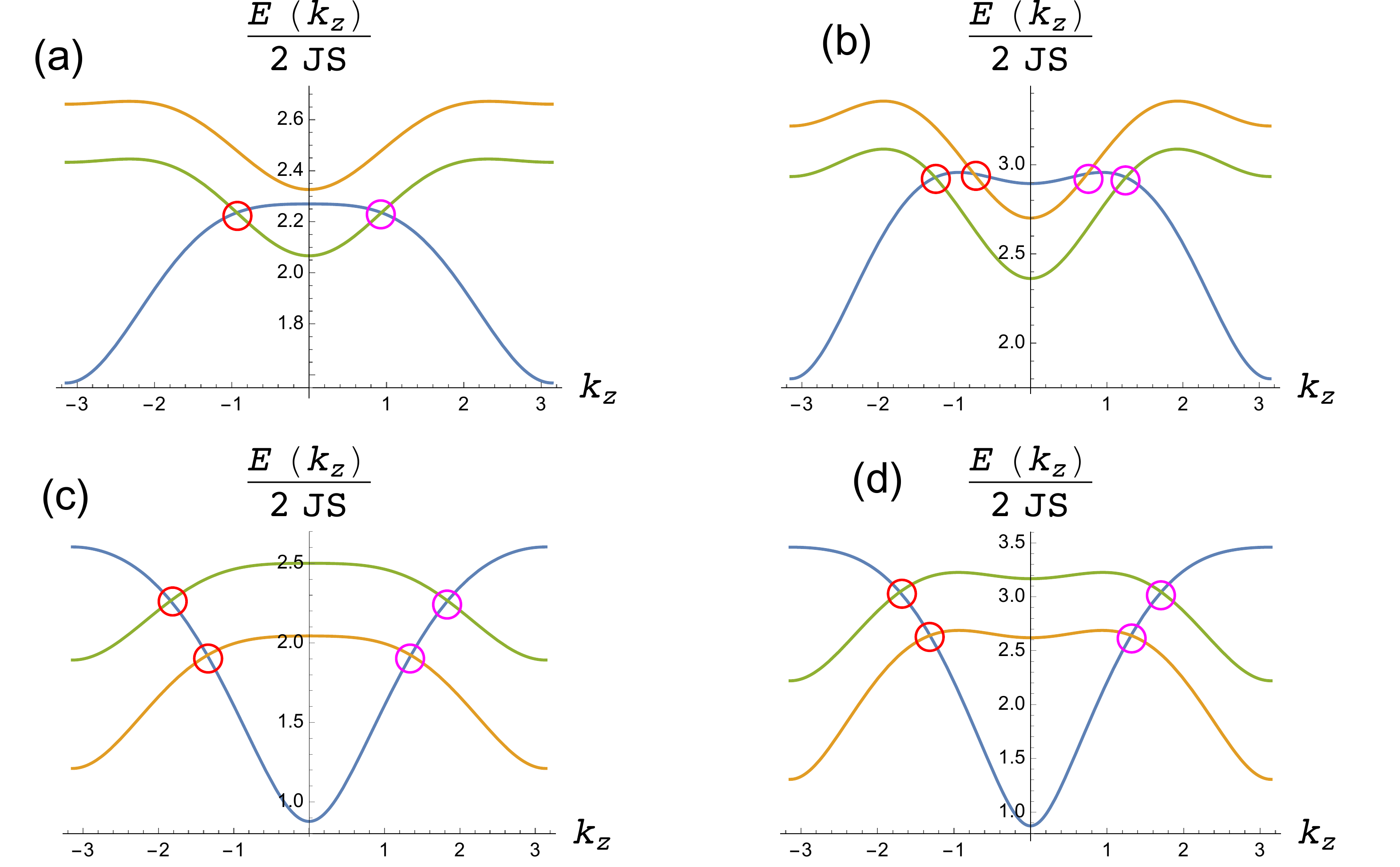}
\caption{Color online.  Weyl  magnon bands along the $k_z$ direction for  $D_z/J=0.3$, $D_p/J=0.5$.  Top panel  $(k_x, k_y)=(2\pi/3, 0)$.  (a). $J_c/J=0.8$  (b). $J_c/J=1.5$. Bottom panel  $(k_x, k_y)=(0, 0)$.  (c). $J_c/J=0.8$  (d). $J_c/J=1.5$.  The WM nodes are highlighted with red and pink circles.}
\label{band}
\end{figure*}
  
 \begin{align}
E_{\text{cl}}&= -3NS^2\Big[ J\lb 1 -3\sin^2\eta\rb +\sqrt{3}D_z\cos^2\eta \nonumber\\&+\sqrt{3}D_p\sin 2\eta+J_c\cos 2\eta \Big],
\label{cene}
\end{align}
where $N$ is the number of sites per unit cell, $S$ is the spin value, and $\eta$ is the canting angle induced by $D_p$. The classical energy is minimized by  
\begin{align}
\tan2\eta = \frac{2\sqrt{3}D_p}{3J+\sqrt{3}D_z+2J_c}.
\label{can}
\end{align}
Evidently, the in-plane DM interaction provide similar spin canting as an external magnetic field \cite{sol}, and leads to noncoplanar chiral spin structure with nonzero scalar spin chirality (see Appendix). For decoupled kagom\'e planes {i.e.}  $J_c=0$, we recover the canting angle of single layer kagom\'e lattice in Ref.~\cite{men1}. 

\section{Topological Weyl magnons}

\subsection{Weyl magnon bands}

  We now study the WM bands in the current model. As we pointed out above, at zero magnetic field the in-plane DM interaction induces similar spin canting  (scalar spin chirality) a nonzero magnetic field induces at zero in-plane DM interaction \cite{sol}. The major difference here is that the in-plane DM interaction is intrinsic to kagom\'e materials and breaks rotational symmetry. Hence, the current model  have gapped modes at $\bo=0$.  
  
  In order to study the magnon bands in this model,  we generalize linear spin wave theory of 2D kagom\'e-lattice antiferromagnets  with in-plane and out-of-plane DM interaction \cite{yil,ono} to 3D unshifted stacked  kagom\'e-lattice antiferromagnets. The basic procedures are outlined  in the Appendix. At  $D_p=0$ and $D_z\neq 0$ the ground state of the Hamiltonian \eqref{mod} is a 3D conventional non-collinear $120^\circ$  magnetic structure with zero scalar spin chirality. Hence, $\mathcal T$ symmetry is broken only by the magnetic order but not macroscopically.  In this limit degenerate magnon bands  form nodal-line magnons (NLMs) as previously shown \cite{sol}.  A nonzero $D_p\neq 0$  induces 3D noncoplanar chiral spin texture  with non-zero scalar spin chirality which breaks $\mathcal T$ symmetry macroscopically. 
  
  Similar to the magnetic field model \cite{sol},  the degeneracy of the magnon bands are lifted and the NLMs are gapped  everywhere except at the WM nodes located at $(\pm 2\pi/3,0,\pm k_0^{1})$ and $(0,0,\pm k_0^{2})$. At $(k_x,k_y)=(\pm 2\pi/3,0)$ and $(k_x,k_y)=(0,0)$ the eigenvalues of the Bogoliubov spin wave Hamiltonian can be found exactly as a function of $k_z$, hence the locations of the WM nodes $k_0^{1(2)}$ (see Appendix). We have shown the WM bands along the $k_z$ directions in Fig.~\eqref{band} in the weakly coupled regime $J_c<J$ and  strongly coupled regime $J_c\geq J$. A distinguishing feature of this model is that without the lowest magnon band the middle and topmost magnon band never cross each other. Therefore the WM nodes comes from the lowest excitations.  In the weakly coupled regime $J_c<J$ the WM nodes at $(\pm 2\pi/3,0,\pm k_0^{1})$ are more dominant than those at $(0,0,\pm k_0^{2})$, whereas in the strongly coupled regime  $J_c\geq J$  those at $(0,0,\pm k_0^{2})$ are more dominant than those at $(\pm 2\pi/3,0,\pm k_0^{1})$ (see the chiral magnon surface states below). Also note that  WM nodes exist for any nonzero  $D_z$ and $D_p$, with a non-negligible interlayer antiferromagnetic interaction $J_c$.   Due to broken  rotational invariance induced by  $D_p\neq 0$  the lowest mode is gapped at $\bo=0$.

\subsection{Monopoles of the Berry curvatures}
The Berry curvature encodes the interesting properties of WMs as well as the anomalous thermal Hall effect.  The Berry curvature is defined from the paraunitary operator that diagonalizes the magnon Bogoliubov Hamiltonian $\mathcal{H}_B(\bo)$ (see Appendices).  For a given magnon band $n$, it can be defined  as
\begin{align}
\Omega_{n, \alpha\beta}^{\gamma}(\bo)=-\sum_{m\neq n}\frac{2\text{Im}[ \braket{\mathcal{P}_{\bo n}|\hat v_\alpha|\mathcal{P}_{\bo m}}\braket{\mathcal{P}_{\bo m}|\hat v_\beta|\mathcal{P}_{\bo n}}]}{\lb E_{n}(\bo)-E_{m}(\bo)\rb^2},
\label{chern2}
\end{align}
where $\hat v_\alpha=\partial \mathcal{H}_B(\bo)/\partial k_\alpha$ defines the velocity operators and $\alpha,\beta,\gamma=x,y,z$, $\mathcal{P}_{\bo n}$ are the paraunitary  operators (eigenvectors) that diagonalize $\mathcal{H}_B(\bo)$  and  $E_{n}(\bo)$ are the eigenvalues. Note that the Berry curvature is a 3-pseudo-vector pointing along  the $\gamma$ directions perpendicular to both the $\alpha$ and $\beta$ directions. The WM nodes come in pairs of opposite chirality, and act as source (monopole) and sink (anti-monopole) of the Berry curvature as shown in Fig.~\eqref{BC}. 
 
\subsection{Topological thermal Hall effect}
   Now, we turn to the main purpose of this paper --- the topological (anomalous) thermal Hall effect due to WMs. As we mentioned above,  the topological or anomalous thermal Hall effect induced by WMs has not been studied in any literature both theoretically and experimentally. We will provide a theoretical description in this section.  The  thermal Hall effect is due to flow of heat current $J_\alpha^\gamma$ under the influence of a thermal gradient  $\nabla_{\beta}T$. It can be derived from linear response theory \cite{alex2,shin1}. The total intrinsic anomalous thermal Hall conductivity can be written as $\kappa_{H}=\lb \kappa_{yz}^x + \kappa_{zx}^y +\kappa_{xy}^z \rb/3$, where
the components $\kappa_{\alpha\beta}^\gamma=-J_\alpha^\gamma/\nabla_{\beta}T$  are  given explicitly by 
\begin{align}
\kappa_{\alpha\beta}^\gamma=- T\int_{{BZ}} \frac{\text{d}\bo}{(2\pi)^3}~ \sum_{n=1}^N c_2\lb f_n^B\rb\Omega_{n, \alpha\beta}^{\gamma}(\bo),
\label{thm}
\end{align}
where   $ f_n^B=\big( e^{E_{n}(\bo)/T}-1\big)^{-1}$ is the Bose function  with the Boltzmann constant set to unity,  and $ c_2(x)=(1+x)\lb \ln \frac{1+x}{x}\rb^2-(\ln x)^2-2\text{Li}_2(-x)$, with $\text{Li}_2(x)$ being the  dilogarithm. Evidently, the anomalous thermal Hall conductivity is the integration of the Berry curvature over the BZ, weighed by the $c_2$ function. Due to the Berry curvature, $\kappa_{\alpha\beta}^\gamma$ is also a 3-pseudo-vector pointing along  the $\gamma$ directions perpendicular to both the $\alpha$ and $\beta$ directions. For $D_p=0$ and $D_z\neq 0$ the 3D conventional non-collinear $120^\circ$  magnetic structure has an effective $\mathcal T$ symmetry defined as the combination of $\mathcal T$ symmetry  and $180^\circ$ spin rotation of the in-plane spin order about the $z$ direction ($\mathcal R_z$), or mirror reflection symmetry of the kagom\'e plane about the  $y$ direction ($\mathcal M_y $). Thus, $\Omega_{n, \alpha\beta}^{\gamma}(\bo)=-\Omega_{n, \alpha\beta}^{\gamma}(-\bo)$ and $\kappa_{\alpha\beta}^\gamma=0$. For $D_p\neq 0$ these symmetries are broken leading to noncoplanar chiral spin texture. Therefore $\kappa_{\alpha\beta}^\gamma$ is expected to be nonzero. As the Berry curvature is maximum near the WM nodes, evidently the major contribution to $\kappa_{\alpha\beta}^\gamma$ comes from these nodes. In addition, the presence of  Bose function suggests that the lowest magnon band and associated Berry curvature have the dominant contribution to $\kappa_{\alpha\beta}^\gamma$ at low temperature when few magnons are thermally excited. This implies that the most important  WM nodes contributing to $\kappa_{\alpha\beta}^\gamma$ must come from the lowest excitation \cite{foot}.
 
 \begin{figure}
\includegraphics[width=1\linewidth]{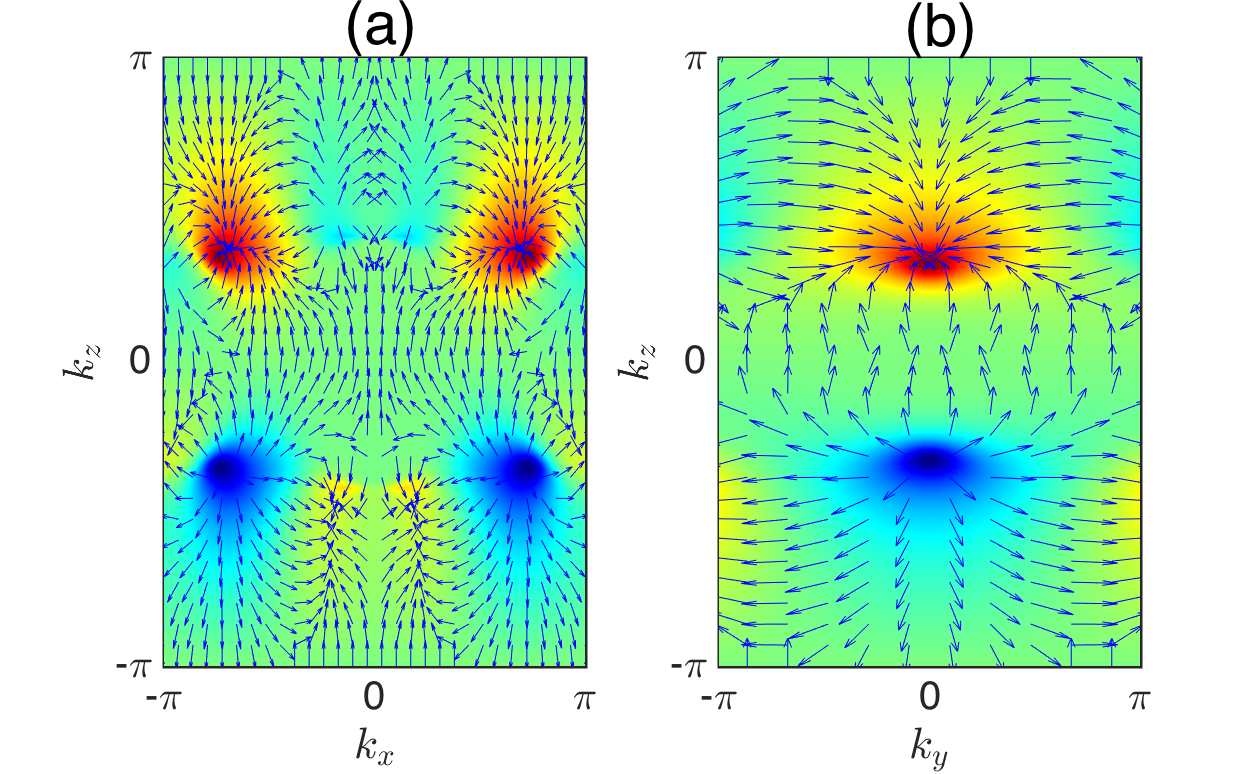}
\caption{Color online. Monopole and anti-monopole distributions of the lowest Weyl magnon band Berry curvature in the (a) $k_y=0$ plane for $\Omega_{1,xz}^y(\vec k)$ and (b) $k_x=2\pi/3$ plane for $\Omega_{1,yz}^x(\vec k)$. The parameters are set to $D_z/J=0.3$, $D_p/J=0.5$, $J_c/J=0.8$ as in Fig.~\ref{band}(a).}
\label{BC}
\end{figure}

\begin{figure}
\includegraphics[width=1\linewidth]{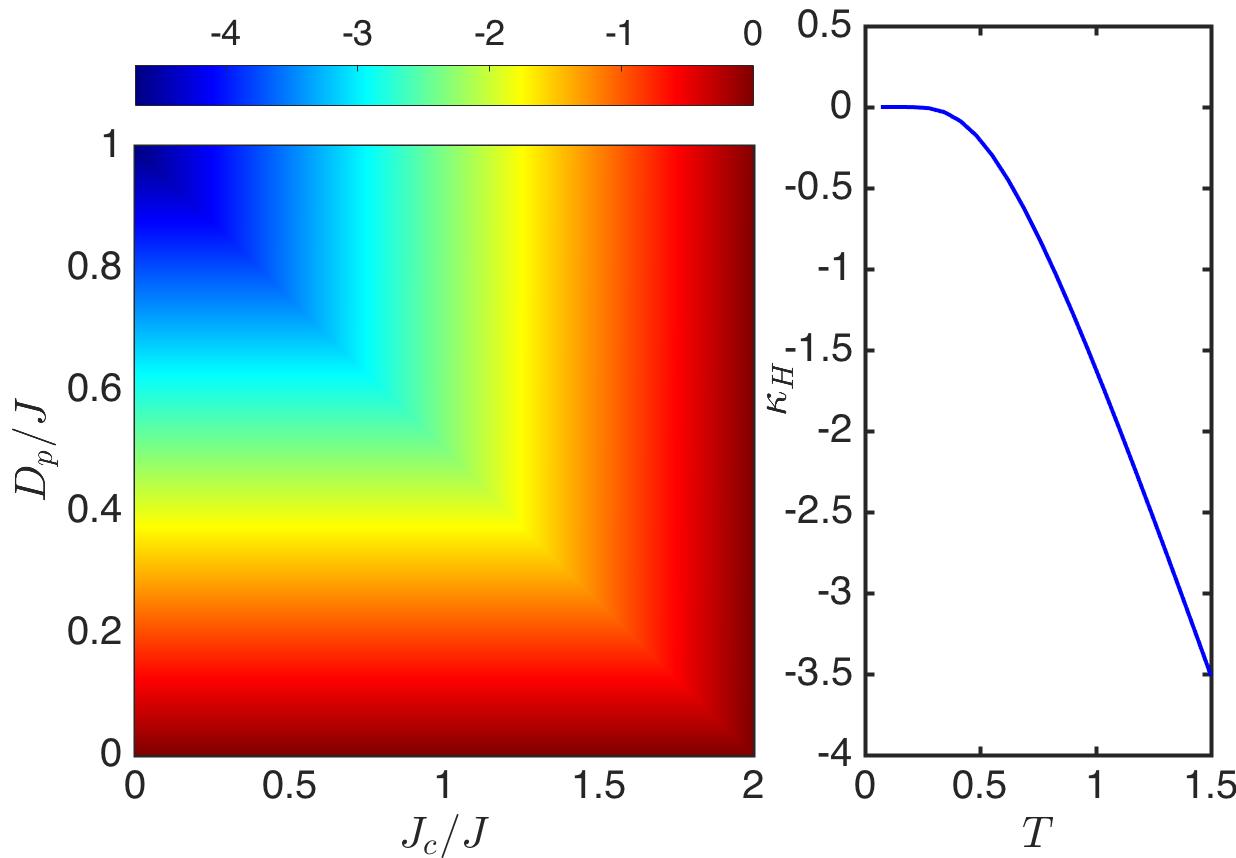}
\caption{Color online.  Topological (anomalous) thermal Hall conductivity. (Left) Heat map of $\kappa_H\approx \kappa_{xy}^z$ in the $J_c/J$--$D_p/J$ plane at $D_z/J=0.3$, $T=0.75$. (Right) $\kappa_H\approx \kappa_{xy}^z$ plotted against $T$ for $D_z/J=0.3$, $D_p/J=0.5$, $J_c/J=0.8$ as in Fig.~\ref{band}(a).}
\label{ATHE}
\end{figure}

For any surface not perpendicular to the $k_z$ direction the WM nodes come in pairs of opposite chirality as shown in Fig.~\eqref{BC}. Therefore, the net Berry curvature vanishes identically.  In  fact, a direct numerical integration of Eq.~\eqref{thm} shows that  $\kappa_{yz}^x = \kappa_{zx}^y\approx 0$. However, for the surface perpendicular to the $k_z$ direction the distributions of WM nodes are different. For any  fixed $k_z$ in the vicinity of the  WM nodes the  net Berry curvature in the $k_x$-$k_y$ plane is nonzero (see Fig.~\eqref{BC}). Therefore, integrating over $k_z$  results in nonzero $\kappa_{xy}^z$ as shown in Fig.~\eqref{ATHE}. The anomalous thermal Hall conductivity vanishes at zero temperature as no magnon is thermal excited. It also vanishes at $D_p=0$ as the scalar spin chirality vanishes, hence $\mathcal T$ symmetry is not broken macroscopically.  The sign of the anomalous thermal Hall conductivity can be easily switched by a small magnetic field.  The  anomalous thermal Hall conductivity $\kappa_{xy}^z$ can be separated in two forms as given by Eqs.~\eqref{ATHC} and \eqref{ATHC1},  from which we infer that empirical formula is given by Eq.~\eqref{sep}  at $T\neq 0$, similar to anomalous Hall conductivity in the Weyl semimetal \cite{ya,bur}. As the WM nodes annihilate at the BZ boundary, the system becomes a fully gapped 3D topological magnon insulator with also nonzero $\kappa_{xy}^z$, similar to 2D system \cite{sol1}. 
 
\section{Conclusion}

We have shown that Weyl magnons in  stacked noncoplanar frustrated kagom\'e antiferromagnets possessed a nonzero intrinsic topological (anomalous) thermal Hall effect at zero magnetic field. It is shown that the topological (anomalous) thermal Hall effect depends on the distributions of the Weyl magnon nodes and it is proportional to their distance  in momentum space at nonzero temperature. We also showed that the Weyl magnon nodes at the lowest excitation carry the dominant contribution to the topological (anomalous) thermal Hall conductivity at low temperature.  Therefore, we  have concluded that the most important Weyl magnons in three-dimensional magnetic systems are those at the lowest excitation \cite{foot}, and they can be easily probed by neutron scattering experiments.  The sign of the topological (anomalous) thermal Hall conductivity can be switched by a small external magnetic field, paving the way toward possible applications in magnon  spintronics and magnetic data storage devices. Moreover, a nonzero topological (anomalous) thermal Hall conductivity in three-dimensional  frustrated kagom\'e antiferromagnets could also serve as an avenue to probe  macroscopically broken time-reversal symmetry or scalar spin chirality. The predicted results can be investigated experimentally  by thermal transport measurements.

\appendix

\section{Spin transformation}
  To facilitate spin wave theory  we express the spins in terms of local axes, such that the $z$-axis coincides with the spin direction \cite{yil,ono}.  The transformation matrix is given by
\begin{align}
\mathcal{R}_t(\theta_{i,\ell})
=\begin{pmatrix}
0 & \sin\theta_{i,\ell} & -\cos\theta_{i,\ell}\\
0 & \cos\theta_{i,\ell} &\sin\theta_{i,\ell}\\
1 & 0 &0
\end{pmatrix},
\end{align}
where $\theta_{i,\ell}$ are the angles that form $120^\circ$ non-collinear spin configuration on sublattice $(1,2,3)$ depicted in  Fig.~\eqref{KL}. Due to spin canting induced by $D_p$ we have to transform the spin from local axes to canting frame using the rotation matrix 
\begin{align}
\mathcal{R}_c(\eta)
=\begin{pmatrix}
\cos\eta &0 & -\sin\eta\\
0 & 1 &0\\
\sin\eta & 0 &\cos\eta
\end{pmatrix}.
\end{align}
 Now, the spins transform as $ {\vec S}_i=\mathcal{R}_z(\theta_{i,\ell})\cdot\mathcal{R}_y(\eta)\cdot{\vec S}_i^\prime,$ where prime denotes the rotated frame. Note that the triangular plaquettes on the kagom\'e lattice have $\mathcal C_3$ rotational symmetry, hence the bonds $1\to 2$, $2\to 3$, and $3\to 1$ in Fig.~\eqref{KL} have   the same coefficients. For instance, for bond $1 \to 2$ the terms that contribute to noninteracting  magnon are  given by

  \begin{align}
  \mathcal H_J^{1\to 2}&= J\sum_{\ell}\Big[-\frac{1}{2} {\vec S}_{1,\ell}^\prime\cdot {\vec S}_{2,\ell}^\prime -\frac{\sqrt{3}}{2}\sin\eta~\hat{z}\cdot\lb {\vec S}_{1,\ell}^\prime\times{\vec S}_{2,\ell}^\prime\rb \nonumber\\& +\frac{3}{2}\lb\cos^2\eta  S_{1,\ell}^{\prime x}S_{2,\ell}^{\prime x} +\sin^2\eta S_{1,\ell}^{\prime z} S_{2,\ell}^{\prime z}\rb\Big],\\
  \mathcal   H_{D_z}^{1\to 2}&=D_z\sum_{\ell}\Big[\frac{1}{2}\sin\eta~  \hat {z}\cdot \lb {\vec S}_{1,\ell}^\prime\times{\vec S}_{2,\ell}^\prime\rb\nonumber\\&-\frac{\sqrt{3}}{2}\lb \sin^2\eta S_{1,\ell}^{\prime x}S_{2,\ell}^{\prime x} + S_{1,\ell}^{\prime y}S_{2,\ell}^{ \prime y}+\cos^2\eta S_{1,\ell}^{\prime z}S_{2,\ell}^{\prime z}\rb\Big],\\
  \mathcal   H_{D_p}^{1\to 2}&= D_p\sum_{\ell}\Big[\frac{\sqrt{3}\sin 2\eta}{2} (S_{1,\ell}^{\prime x}S_{2,\ell}^{\prime x} -S_{1,\ell}^{\prime z}S_{2,\ell}^{\prime z}) \nonumber\\&+\frac{\cos\eta}{2}~\hat {z}\cdot \lb {\vec S}_{1,\ell}^\prime\times{\vec S}_{2,\ell}^\prime\rb \Big].
  \label{rota}
  \end{align}
The  $\mathcal C_3$ rotational symmetry guarantees that  bonds $2\to 3$ and $3\to 1$ have the same form of Hamiltonians. The interlayer coupling transforms as
 \begin{align}
    \mathcal H_{J_c}&= J_c\sum_{ i,\la \ell\ell^\prime\ra}\Big[\cos\theta_{\ell\ell^\prime} {\vec S}_{i,\ell}^\prime\cdot {\vec S}_{i,\ell^\prime}^\prime \nonumber\\&+2\sin^2\lb\frac{\theta_{\ell\ell^\prime}}{2}\rb\lb\cos^2\eta  S_{i,\ell}^{\prime x}S_{i,\ell^\prime}^{\prime x} +\sin^2\eta S_{i,\ell}^{\prime z} S_{i,\ell^\prime}^{\prime z}\rb\Big],
  \label{rotc}
  \end{align}
where $\theta_{\ell\ell^\prime}=\theta_{\ell}-\theta_{\ell^\prime}$. Here $\theta_{\ell\ell^\prime}=\pi$ for antiferromagnetic interlayer coupling $J_c>0$, and   $\theta_{\ell\ell^\prime}=0$ for ferromagnetic interlayer $J_c<0$. Note that the scalar spin chirality of the noncoplanar  (umbrella) spin configurations defined as 
\bea
\chi= \sum_{ijk,l} {\vec S}_{i,\ell}^\prime\cdot\lb {\vec S}_{j,\ell}^\prime\times{\vec S}_{k,\ell}^\prime\rb,
\eea  is induced only within the kagom\'e  planes. It vanishes at $D_p=0$ by virtue of   the classical ground state energy and the canting angle given by Eqs.~\eqref{cene} and \eqref{can} in the main text.
 
\section{Holstein-Primakoff transformation}
 Next, we introduce the  Holstein-Primakoff  bosons: 
\begin{align}
S_{i,\ell}^{z}= S-a_{i,\ell}^\dagger a_{i,\ell},~  S_{i,\ell}^{+} \approx  \sqrt{2S}a_{i,\ell}=(S_{i,\ell}^{-})^\dg,
\end{align} where $ S_{i,\ell}^{\pm}=S_{i,\ell}^{x}\pm i S_{i,\ell}^{y}$ and $a_{i,\ell}^\dagger(a_{i,\ell})$ are the bosonic creation (annihilation) operators.  The resulting magnon tight-binding model is given by

\begin{align}
\mathcal H_{J-D_{z(p)}}&= S\sum_{\la ij\ra,\ell}\big[ t^z(a_{i,\ell}^\dg a_{i,\ell} +a_{j,\ell}^\dg a_{j,\ell})\nonumber\\&+ t^r(e^{-i\phi_{ij,\ell}}a_{i,\ell}^\dg a_{j,\ell} + h.c.)+ t^o(a_{i,\ell}^\dg a_{j,\ell}^\dg + h.c.)\big],\\
\mathcal H_{J_c}&= S\sum_{i,\ell} t_c^za_{i,\ell}^\dg a_{i,\ell}+ S\sum_{i,\la \ell\ell^\prime\ra}\big[ t_c^r(a_{i,\ell}^\dg a_{i,\ell^\prime} + h.c.)\nonumber\\& + t_c^o(a_{i,\ell}^\dg a_{i,\ell^\prime}^\dg + h.c.)\big].
\end{align}
The parameters of the tight binding model are given by 
\begin{align}
& t^z= \frac{J}{2}\lb 1 - 3\sin^2\eta\rb +\frac{\sqrt{3}}{2}D_z\cos^2\eta +\frac{\sqrt{3}D_p}{2}\sin 2\eta,\\&
t^r=\sqrt{(t^{r}_1)^2+(t^{r}_2)^2},\\
&t_1^{r}=\frac{J}{2}\Big[-1+\frac{3}{2}\cos^2\eta\Big]-\frac{\sqrt{3}D_z}{2}\lb 1-\frac{\cos^2\eta}{2}\rb\nonumber\\&+\frac{\sqrt{3}D_p}{4}\sin 2\eta,\\
&t_2^{r}= -\frac{1}{2}\big[( \sqrt{3}J -D_z)\sin\eta-D_p\cos\eta\big],\\
&t^o=\frac{1}{4}(3J+\sqrt{3}D_z)\cos^2\eta +\frac{\sqrt{3}D_p}{4}\sin 2\eta,~\\
& t_c^z= 2J_{c}\cos 2\eta,~
t_c^r=-  J_{c}\sin^2\eta,~
t_c^o=J_{c} \cos^2\eta.
\end{align}
 The  solid angle subtended by three noncoplanar spins  is given by $\phi_{ij}=\pm\phi$, where $\phi=\tan^{-1}[t_2^{r}/t_1^{r}]$. Next, we Fourier transform into momentum space with   basis vector $\psi^\dg_\bo=(a_{\bo 1}^{\dg},\thinspace a_{\bo 2}^{\dg},\thinspace a_{\bo 3}^{\dg}, \thinspace a_{-\bo 1},\thinspace a_{-\bo 2},\thinspace a_{-\bo 3} )$. The resulting  Hamiltonian  is given by

 \begin{align}
& \mathcal{H}(\bo)= 2S\begin{pmatrix}
  { \Lambda^{0}}(k_z)+ \Lambda^r(\vec k_\parallel)&  \Lambda^o(\vec k_\parallel,k_z)\\
 \Lambda^o(\vec k_\parallel,k_z) &\  { \Lambda^{0}}(k_z)+ \Lambda^r(\vec k_\parallel)
\end{pmatrix},
\label{eqnr}
\end{align}
where $\bo=(k_\parallel,k_z)$ and $\vec k_\parallel=(k_x,k_y)$. The  $ \Lambda$ matrices are given by $ \Lambda^{0}(k_z)=[(t_c^z +4t^z)/2+t_c^r\cos k_z]{I}_{3\times 3}$
\begin{align}
 \Lambda^{r}(k_\parallel)= t^r
\begin{pmatrix}
0& \cos k_\parallel^1e^{-i\phi}& \cos k_\parallel^3 e^{i\phi}\\
\cos k_\parallel^1e^{i\phi}&0&\cos k_\parallel^2e^{-i\phi}\\
\cos k_\parallel^3e^{-i\phi}&\cos k_\parallel^2e^{i\phi}&0
\end{pmatrix},
\label{mat1}
\end{align}
\begin{align}
 \Lambda^{o}(k_\parallel,k_z)= 
\begin{pmatrix}
t_c^o\cos k_z& t^o\cos k_\parallel^1& t^o\cos k_\parallel^3 \\
t^o\cos k_\parallel^1&t_c^o\cos k_z &t^o\cos k_\parallel^2\\
t^o\cos k_\parallel^3&t^o\cos k_\parallel^2&t_c^o\cos k_z
\end{pmatrix},
\label{mat2}
\end{align}
where  $k_\parallel^i=\vec k_\parallel\cdot{\vec  a}_i$, with   ${\vec a}_1={\hat x}$, ${\vec a}_2={\hat x}/2+\sqrt{3}\hat y/2$, and ${\vec a}_3=-{\hat x}/2+\sqrt{3}\hat y/2$. The momentum space Hamiltonian for $J_c<0$ can be derived in a similar way. However, we will concentrate on antiferromagnetic interlayer coupling  $J_c>0$ as it possesses Weyl magnon nodes.
The Hamiltonian is diagonalized through generalized Bogoliubov transformation. This follows by making  a linear  transformation 
$\psi_\bo= \mathcal{P}_\bo Q_\bo$, 
where   $Q^\dg_\bo= (b_{\bo 1}^{\dg},\thinspace b_{\bo 2}^{\dg},\thinspace b_{\bo 3}^{\dg},b_{-\bo 1},\thinspace b_{-\bo 2},\thinspace b_{-\bo 3})$ is the quasiparticle operators, and $\mathcal{P}_\bo$ is a $2N\times 2N$ paraunitary matrix  defined as
\begin{align}
& \mathcal{P}_\bo= \begin{pmatrix}
  u_\bo& -v_\bo^* \\
-v_\bo&u_\bo^*\\  
 \end{pmatrix}.
\end{align} 
The functions $u_\bo$ and $v_\bo$ are  $N\times N$ matrices and they satisfy the relation \bea |u_\bo|^2-|v_\bo|^2={I}_{N\times N}.\eea  Whereas the paraunitary operator $\mathcal{P}_\bo$ satisfies the relations,
\begin{align}
&\mathcal{P}_\bo^\dg \mathcal{H}(\bo) \mathcal{P}_\bo=\mathcal{E}_\bo,\label{eqn1}\\ &\mathcal{P}_\bo^\dg  {\tau}_3 \mathcal{P}_\bo=  {\tau}_3,
\label{eqna}
\end{align}
where $\mathcal{E}_\bo=\text{diag}(E_{\bo n},E_{-\bo n})$,~${\tau}_3=\text{diag}({I}_{N\times N}, -{I}_{N\times N} )$, $E_{\bo,n}$ are the  eigenmodes for band $n$, and ``\text{diag}'' denotes diagonal matrix. From Eq.~\eqref{eqna} the relation $\mathcal{P}_\bo^\dg= {\tau}_3 \mathcal{P}_\bo^{-1} {\tau}_3$ holds. Therefore, from Eq.~\eqref{eqn1} the Hamiltonian to be diagonalized is given by $\mathcal{H}_B(\bo)= {\tau}_3\mathcal{H}(\bo)$, whose eigenvalues are given by $ {\tau}_3\mathcal E_\bo$ and the columns of $\mathcal P_\bo$ are the corresponding eigenvectors. For $D_p=0$ the system has U(1) rotational invariance and degenerate energy bands form nodal-line magnons (NLMs) as we have previously discussed \cite{sol}.

\section{Weyl magnon bands}
 For $D_p\neq 0$ rotational invariance and time-reversal symmetry are explicitly broken due to noncoplanar chiral spin textures. Therefore the possibility of Weyl magnons (WMs) becomes possible as discussed in the main text.  At $(k_x,k_y)=(\pm 2\pi/3, 0)$ and $(k_x,k_y)=(0, 0)$ the eigenvalues of $\mathcal{H}_B(\bo)$ can be found exactly as a function of $k_z$, hence the location of the WM nodes. 

At $(k_x,k_y)=(\pm 2\pi/3, 0)$, the magnon bands are given by
\begin{align}
[E_0(k_z)]^2&=\frac{1}{2}\Big[2\lbrace\mathcal{G}^0(k_z)\rbrace^2 + (t^r)^2-(t_c^o)^2-2(t^o)^2\nonumber\\&+4t_c^ot^o\cos(k_z)-(t_c^o)^2\cos(2k_z)\\&\nonumber-4t^r\mathcal{G}^0(k_z)\cos(\phi)+(t^r)^2\cos(2\phi)\Big].
\end{align}
\begin{align}
[E_\pm(k_z)]^2&=\frac{1}{2}\Big[\lbrace 2\mathcal{G}^0(k_z)\rbrace^2 + 2(t^r)^2-2(t_c^o)^2-(t^o)^2\nonumber\\&-2t_c^o\lbrace 2t^o\cos(k_z)+ 4t_c^o\cos(2k_z)\rbrace\\&\nonumber+4t^r\mathcal{G}^0(k_z)\cos(\phi)-(t^r)^2\cos(2\phi)
\nonumber\\&\pm 2\sqrt{3} t^r\sin(\phi)[2\mathcal{G}^0(k_z)+t^r\cos(\phi)]\Big],
\end{align}
where subscript $0$ denotes lowest band, whereas $\mp$ denotes middle  and topmost bands respectively.

At $(k_x,k_y)=(0, 0)$, the magnon bands are given by
\begin{align}
[E_0(k_z)]^2&=\frac{1}{2}\Big[2\lbrace \mathcal{G}^0(k_z)\rbrace^2 + (2t^r)^2-(t_c^o)^2-2(2t^o)^2\nonumber\\&-t_c^o\lbrace 8t^o\cos(k_z)+t_c^o\cos(2k_z)\rbrace\\&\nonumber+8t^r\mathcal{G}^0(k_z)\cos(\phi)+(2t^r)^2\cos(2\phi)\Big].
\end{align}
\begin{align}
[E_\pm(k_z)]^2&=\frac{1}{2}\Big[2\lbrace \mathcal{G}^0(k_z)\rbrace^2 + (2t^r)^2-(t_c^o)^2-2(t^o)^2\nonumber\\&+t_c^o\lbrace 4t^o\cos(k_z)- t_c^o\cos(2k_z)\rbrace\\&\nonumber-4t^r\mathcal{G}^0(k_z)\cos(\phi)-2(t^r)^2\cos(2\phi)
\nonumber\\&\pm 4\sqrt{3} t^r\sin(\phi)[\mathcal{G}^0(k_z)-t^r\cos(\phi)]\Big].
\end{align}

The lowest  and middle magnon  bands cross linearly (form WM nodes) at $(\pm 2\pi/3, 0, k_0^{1})$  and $(0, 0, k_0^{2})$, where 
\begin{align}
k_0^{1}&=\pm \cos^{-1}(\alpha_1/\beta_1)\\
 k_0^{2}&=\pm \cos^{-1}(\alpha_2/\beta_2),
\end{align}

\begin{align}
\alpha_1 &= 3(t^o)^2 + t^r\Big[-3t^r\cos(2\phi)+ 6\cos(\phi)\lbrace -1+\sqrt{3}D_z\nonumber\\&+(3+\sqrt{3}D_z+2Jc)\cos(2\eta) +2\sqrt{3}D_p\sin(2\eta)\rbrace\nonumber\\&-2\lbrace -\sqrt{3}+3D_z+[3D_z+\sqrt{3}(3+2J_c)]\cos(2\eta)\nonumber\\&+\sqrt{3}t^r\cos(\phi)+6D_p\sin(2\eta)\rbrace\sin(\phi)\Big],\\
\beta_1 &= 12t_c^ot^o-4t^rt_c^r\big[3\cos(\phi)-\sqrt{3}\sin(\phi)\big].
\end{align}

\begin{align}
\alpha_2 &= -3(t^o)^2 + t^r\Big[ 3t^r\cos(2\phi)+3\cos(\phi)\lbrace -1+\sqrt{3}D_z\nonumber\\&+(3+\sqrt{3}D_z+2J_c)\cos(2\eta)+ 2\sqrt{3}D_p\sin(2\eta)\rbrace \nonumber\\& +\lbrace 6D_z\cos^2(\eta)+\sqrt{3}(3+2J_c)\cos(2\eta)\nonumber\\&-\sqrt{3}(1+2t^r\cos(\phi))+6D_p\sin(2\eta)\rbrace\sin(\phi)\Big],\\
\beta_2 &= 6t_c^ot^o-2t^rt_c^r [3\cos(\phi)+\sqrt{3}\sin(\phi)].
\end{align}

The topmost and lowest magnon  bands cross linearly (form WM nodes) at $(\pm 2\pi/3, 0, \tilde k_0^{1})$ and  $(0,0,  \tilde k_0^{2})$, where 
\begin{align}
\tilde k_0^{1}&=\pm \cos^{-1}( \tilde\alpha_1/\tilde \beta_1),\\
\tilde k_0^{2}&=\pm \cos^{-1}(\tilde\alpha_2/\tilde \beta_2),
\end{align}
\begin{align}
\tilde \alpha_1 &= -3(t^o)^2 - t^r\Big[-3t^r\cos(2\phi)+ 6\cos(\phi)\lbrace -1+\sqrt{3}D_z\nonumber\\&+(3+\sqrt{3}D_z+2Jc)\cos(2\eta) +2\sqrt{3}D_p\sin(2\eta)\rbrace\nonumber\\&+2\lbrace -\sqrt{3}+3D_z+[3D_z+\sqrt{3}(3+2J_c)]\cos(2\eta)\nonumber\\&+\sqrt{3}t^r\cos(\phi)+6D_p\sin(2\eta)\rbrace\sin(\phi)\Big],\\
\tilde \beta_1 &= -12t_c^ot^o+4t^rt_c^r\big[\cos(\phi)+\sqrt{3}\sin(\phi)\big],\\
\tilde \alpha_2 &= -3(t^o)^2 + t^r\Big[ 3t^r\cos(2\phi)+3\cos(\phi)\lbrace -1+\sqrt{3}D_z\nonumber\\&+(3+\sqrt{3}D_z+2J_c)\cos(2\eta)+ 2\sqrt{3}D_p\sin(2\eta)\rbrace \nonumber\\& +\lbrace \sqrt{3}-3D_z-[3D_z+\sqrt{3}(3+2J_c)]\cos(2\eta)\nonumber\\&+2\sqrt{3}t^r\cos(\phi)-6D_p\sin(2\eta)\rbrace\sin(\phi)\Big],\\
\tilde \beta_2 &= 6t_c^ot^o-2t^rt_c^r[3 \cos(\phi)-\sqrt{3}\sin(\phi)].
\end{align}


\begin{thebibliography}{99}
\bibitem{foot}
   Similarly, electronic Weyl nodes close to the Fermi energy are the most important ones that make  significant contributions to the low-temperature transport properties.
\end{thebibliography}
\end{document}